\pdfoutput=1

\documentclass[aps,prb,twocolumn,superscriptaddress]{revtex4}
\usepackage{graphicx} %graphics handler
\usepackage{dcolumn}% Align table columns on decimal point
\usepackage{bm}% bold math

\begin{document}

\title{Plasmonic Nature of the Terahertz Conductivity Peak in Single-Wall \\Carbon Nanotubes}
 
\author{Q. Zhang}
\affiliation{Department of Electrical and Computer Engineering, Rice University, Houston, TX 77005, USA}

\author{E. H. H\'{a}roz}
\altaffiliation{Present address: Center for Integrated Nanotechnologies, Los Alamos National Laboratory, Los Alamos, NM 87545, USA}
\affiliation{Department of Electrical and Computer Engineering, Rice University, Houston, TX 77005, USA}

\author{Z. Jin}
\author{L. Ren}
\author{X. Wang}
\affiliation{Department of Electrical and Computer Engineering, Rice University, Houston, TX 77005, USA}

\author{R. S. Arvidson}
\altaffiliation{Present address: MARUM, University of Bremen, Germany}
\affiliation{Department of Earth Science, Rice University, Houston, TX 77005, USA}

\author{A. L\"{u}ttge}
\altaffiliation{Present address: MARUM, University of Bremen, Germany}
\affiliation{Department of Earth Science, Rice University, Houston, TX 77005, USA}
\affiliation{Department of Chemistry, Rice University, Houston, TX 77005, USA}

\author{J.~Kono}
\email[]{kono@rice.edu}
\thanks{corresponding author.}
\affiliation{Department of Electrical and Computer Engineering, Rice University, Houston, TX 77005, USA}
\affiliation{Department of Physics and Astronomy, Rice University, Houston, TX 77005, USA}
\affiliation{Department of Materials Science and NanoEngineering, Rice University, Houston, TX 77005, USA}

\date{\today}

\begin{abstract}
Plasmon resonance is expected to occur in metallic and doped semiconducting carbon nanotubes in the terahertz frequency range, but its convincing identification has so far been elusive.  The origin of the terahertz conductivity peak commonly observed for carbon nanotube ensembles remains controversial.  Here we present results of optical, terahertz, and DC transport measurements on highly enriched metallic and semiconducting nanotube films.  A broad and strong terahertz conductivity peak appears in both types of films, whose behaviors are consistent with the plasmon resonance explanation, firmly ruling out other alternative explanations such as absorption due to curvature-induced gaps.
\end{abstract}

%\pacs{78.67.Ch,71.35.Ji,78.55.-m}
%\keywords{}

\maketitle


\section{Introduction}

%\section{INTRODUCTION}
Understanding the dynamic and plasmonic properties of charge carriers in single-wall carbon nanotubes (SWCNTs) is crucial for emerging applications of SWCNT-based ultrafast electronics and optoelectronics devices,\cite{AvourisetAl08NP,NanotetAl12AM} especially in the terahertz (THz) range.\cite{PortnoietAl08SM,RenetAl12JIMT}  SWCNTs with different chiralities exhibit either semiconducting or metallic properties, providing great flexibility for a variety of THz and plasmonic applications, including sources,\cite{PortnoietAl06SPIE,KibisetAl07NL,NemilentsauetAl07PRL} detectors,\cite{PortnoietAl06SPIE,PortnoietAl08SM} antennas,\cite{Hanson05IEEE,BurkeetAl06IEEE} and polarizers.\cite{RenetAl09NL,KyoungetAl11NL,RenetAl12NL}  A pronounced, finite-frequency peak in THz conductivity spectra has been universally observed in diverse types of SWCNT samples, containing both semiconducting and metallic nanotubes.\cite{BommelietAl96SSC,UgawaetAl99PRB,ItkisetAl02NL,JeonetAl02APL,JeonetAl04JAP,JeonetAl05JAP,AkimaetAl06AM,BorondicsetAl06PRB,NishimuraetAl07APL,KampfrathetAl08PRL,SlepyanetAl10PRB,PekkerKamars11PRB,ShubaetAl12PRB,RenetAl13PRB}  Two interpretations have emerged regarding the THz peak, but there is no consensus about its origin.  One of the possible mechanisms proposed by many authors\cite{UgawaetAl99PRB,BorondicsetAl06PRB,NishimuraetAl07APL,KampfrathetAl08PRL}  is based on interband absorption across the curvature-induced bandgap\cite{Hamadaetal92PRL,KaneMele97PRL} in non-armchair metallic SWCNTs, while the other is the plasmon resonance in metallic and doped semiconducting SWCNTs due to their finite lengths.\cite{JeonetAl02APL,AkimaetAl06AM,SlepyanetAl10PRB,ShubaetAl12PRB,NakanishiAndo09JPSJ}  Hence, spectroscopic studies of type-sorted SWCNT samples are crucial for determining which of the two working hypotheses is correct.

In the first scenario, direct interband absorption occurs across the narrow bandgap\cite{Hamadaetal92PRL,KaneMele97PRL,OuyangetAl01Science}  induced by lattice distortion in the rolled-up graphene sheet in non-armchair metallic SWCNTs.  The magnitude of the induced bandgap is given by\cite{KaneMele97PRL} $E_{\rm g}^{\rm ind}$ = ${3 \gamma_0 a_{\rm C-C}^2 \over 4 d_t^2} \cos{3\alpha}$, where $a_{\rm C-C}$ = 0.142~nm is the interatomic distance in graphene, $d_t$ is the nanotube diameter, $\gamma_0$ $\sim$ 3.2~eV is the tight-binding transfer integral, and $\alpha$ is the chiral angle.  For a SWCNT with $d_t$ $\sim$ 1.5~nm and $\alpha$ $\sim$ 0, $E_{\rm g}^{\rm ind}$ $\sim$ 20~meV.\cite{HarozetAl13NS}  If this scenario is correct, the THz peak should (1)~appear only in non-armchair ($\alpha \ne 30^{\circ}$) metallic SWCNTs, (2)~show a sensitive dependence on $d_t$,\cite{PekkerKamars11PRB} (3)~be suppressed by doping or optical pumping,\cite{KampfrathetAl08PRL} have a strong (4)~temperature dependence and (5)~polarization dependence.  It is also important to note that no detailed theory exists for predicting the lineshape for interband absorption in non-armchair metallic SWCNTs including excitonic effects.\cite{Ando97JPSJ,HartmannetAl11PRB}

In the second scenario, THz radiation incident onto a finite-length metallic or doped semiconducting SWCNT launches collective charge oscillations (plasmons) along the nanotube axis with a frequency given by the charge density and nanotube length.  The expected experimental signatures of this process are: (1)~the THz peak should be observable both in metallic and doped semiconducting nanotubes; (2)~the frequency of the THz peak should systematically depend on the nanotube length in a predictable manner;\cite{NakanishiAndo09JPSJ,ShubaetAl12PRB} (3)~the THz peak intensity should be enhanced by doping in semiconducting SWCNTs; (4)~there should be weak temperature dependence;\cite{UgawaetAl99PRB}  and (5)~there should be strong polarization dependence, with no resonance expected for polarization perpendicular to the nanotube axis.

To systematically prove or disprove these scenarios, broadband spectroscopic studies on well-separated semiconducting and metallic SWCNT samples are necessary.\cite{IchidaetAl11SSC} In particular, in order to correctly interpret THz spectra, it is important to monitor interband transitions in the near-infrared (NIR), visible (VIS), and ultraviolet (UV).  Hence, we performed absorption spectroscopy studies from the THz to the UV as well as DC transport measurements on highly-enriched semiconducting and metallic SWCNT films.  We clearly observed a broad and pronounced THz peak in both types of films.  However, in semiconductor-enriched films, the peak significantly decreased in intensity after carriers were removed by annealing.  In both types of films, the THz peak showed a very weak temperature dependence.  These observations led us to conclude that this peak is due to plasmon resonance of free carriers in metallic and doped semiconducting nanotubes, rather than interband absorption in non-armchair metallic nanotubes.

\section{Methods}
\label{methods}

Suspensions enriched in metallic and semiconducting types of SWCNTs were produced via density gradient ultracentrifugation (DGU)\cite{ArnoldetAl05NL,ArnoldetAl06NN,YanagietAl08APE,HarozetAl10ACS,HarozetAl12JACS,HarozetAl13NS}  using a three-surfactant system.  Arc-discharge-P2-SWNT grade (Carbon Solutions, Inc.)~SWCNTs were initially dispersed in 1\% (wt./vol.) sodium deoxycholate (DOC) (sodium deoxycholate monohydrate, Aldrich, 97\% purity) by bath sonication (Cole-Parmer 60~W ultrasonic cleaner, model \#08849-00) for 30 minutes at a starting concentration of SWCNTs of 1~g/L.  The suspension was then further sonicated by probe ultrasonicator (Cole-Parmer 500~W ultrasonic processor, model \# CPX-600, 1/4" probe, 35\% amplitude) for 6 hours while being cooled in a water bath maintained at 10$^{\circ}$C.  Lastly, the suspension was then centrifuged for 1 hour at 208,400g average (Sorvall Discovery 100SE Ultracentrifuge using a Beckman SW-41 Ti swing bucket rotor) to remove large bundles of SWCNTs.  After centrifugation, the upper 80\% of the supernatant was removed for use in DGU.

For DGU, a mass density gradient was prepared composed of 1.5\% (wt./vol.) sodium dodecyl sulfate (SDS) (sodium dodecyl sulfate- molecular biology or electrophoresis grade, Sigma, 99\% purity), 1.5 \% (wt./vol.) sodium cholate (SC) (sodium cholate hydrate, Aldrich, 98\% purity), and varying amounts of iodixanol (Opti-Prep density gradient medium, Sigma, 60\% (wt./vol.) solution in water).  The gradient was layered inside a centrifuge tube in 2~mL volume steps starting from the bottom with 40\% (wt./vol.), 30\% (wt./vol.), 27.5\% (wt./vol.), 25\% (wt./vol.), 22.5\% (wt./vol.), and 20\% (wt./vol.) iodixanol.  All gradient steps except the 40\% layer contained 1.5\% (wt./vol.) SDS and 1.5\% (wt./vol.) SC.  The 40\% layer, which contained SWCNTs, was prepared by vortex mixing (Fisher Scientific Vortex Genie 2, model \# 12-812, mixed at maximum setting) for 2 minutes 0.67~mL of the SWCNT supernatant prepared previously with 1.33~mL of 60\% iodixanol, 1.5\% SDS, and 1.5\% SC to ultimately form 2~mL of 40\% iodixanol, 1\% SDS, 1\% SC, and 0.33\% DOC.  The gradient was then centrifuged for 18 hours at 208,400g average (Beckman SW-41 Ti swing bucket rotor).  The top 250~$\mu$L of the metal-enriched (aqua green-blue-colored) fractions and semiconductor-enriched (rust-colored) fractions were extracted by hand-pipetting for measurements of suspension samples.  

For the preparation of film samples, vacuum filtration was performed on enriched SWCNT suspensions using a procedure modified from one previously described.\cite{DanetAl12IECR}  Briefly, 8~mL of enriched SWCNT suspension was vacuum-filtered using 50~nm-pore size, polycarbonate filtration membranes.  After filtration was finished and the film was dried, the film was washed by alternating filtrations of 50~mL of boiling nanopure water and 50~mL of 50\%/50\% vol.~ethanol/nanopure water through the SWCNT film and filtration membrane.  This was repeated three times to remove surfactants and iodixanol.  Finally, the films were immersed in bath of N-methyl-2-pyrrolidone (NMP) to dissolve away the filtration membrane and transferred onto intrinsic silicon and sapphire substrates for optical measurements.  The thickness of semiconductor- and metal-enriched films were measured via vertical scanning interferometry to be 410~nm and 190~nm, respectively.  Annealing of prepared nanotube films was done in a vacuum quartz tube furnace at a base pressure of 3 $\times$ 10$^{-6}$~Torr.  The films were initially heated to 423~K, 573~K, and lastly 823~K.  At each temperature, the films remained at that temperature until no other chemical species were observed to desorb as determined by a monitoring furnace pressure.   Upon completion of annealing, the films were brought back to room temperature under vacuum and then subsequently exposed to the atmosphere.

%46. Dan, B.; Ma, A. W. K.; H‡roz, E. H.; Kono, J.; Pasquali, M.; Industrial and Engineering Chemistry Research 2012, 51, 10232-10237.

For obtaining attenuation spectra in a transmission geometry in a wide spectral range, we used time-domain THz spectroscopy (TDTS) in the 0.15-2.5~THz range, Fourier-transform infrared spectroscopy (FTIR, Jasco-660) in the 3-300~THz range, and a UV-VIS-NIR double beam spectrophotometer (Shimadzu UV-3101PC scanning spectrophotometer) in the 300-1200~THz range (250-1000~nm in wavelength).  The TDTS setup used a ZnTe crystal for THz generation and photoconductive antennas for detection with a 150~fs Ti:sapphire laser.  DC transport studies were performed on the same films via four-point measurements with gold ohmic contacts under a nitrogen environment.

\section{Results}

%%%%% Fig. 1 %%%%%
\begin{figure}
\begin{center}
\includegraphics[scale=0.8]{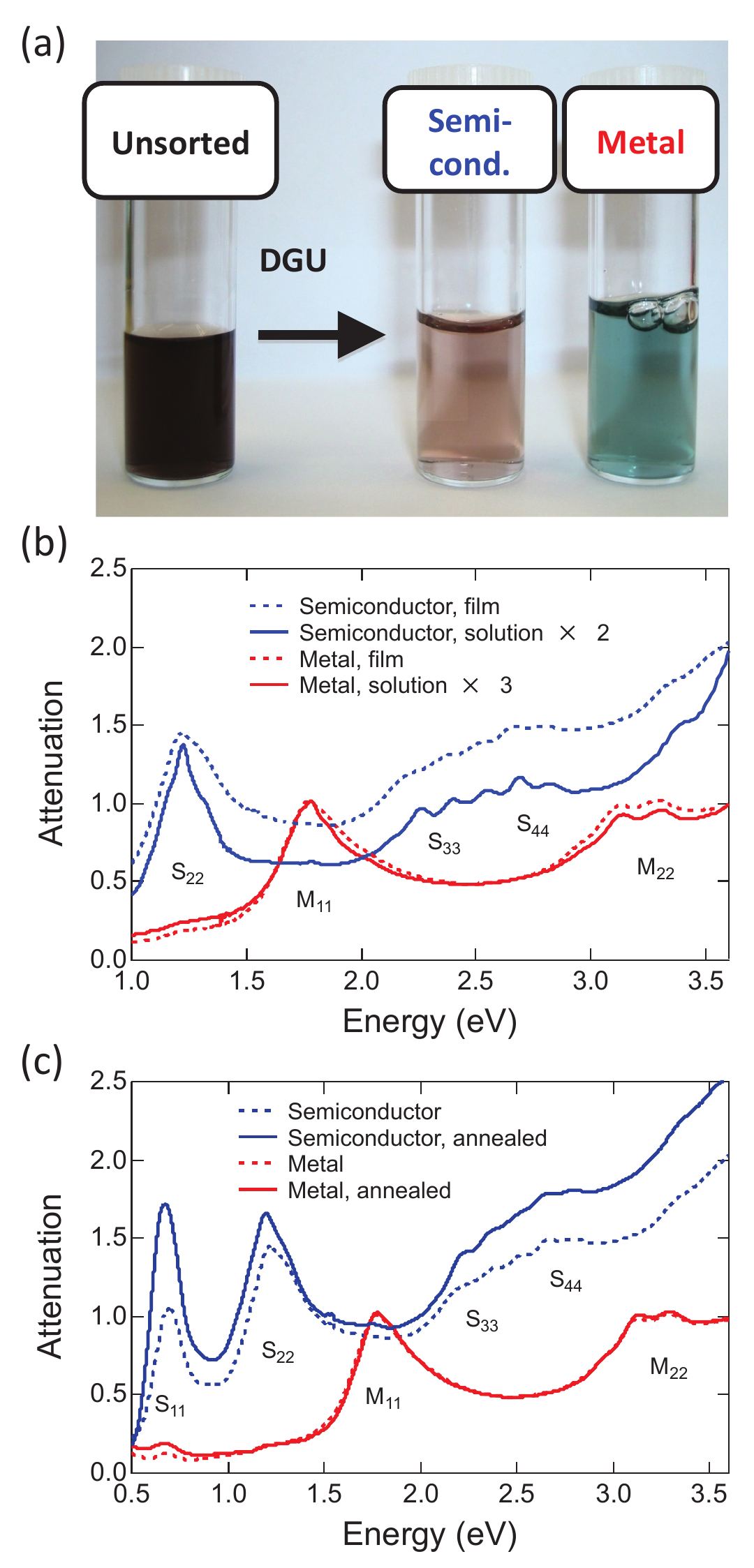}
\caption{(a)~Unsorted, metal-enriched, and semiconductor-enriched SWCNTs suspended in aqueous solution. The type separation is achieved by density gradient ultracentrifugation (DGU). (b)~NIR-to-UV attenuation spectra for semiconductor-enriched (blue curves) and metal-enriched (red curves) SWCNT samples in solution (solid curves) and film (dashed curves) forms. (c)~MIR-to-UV attenuation spectra for semiconductor-enriched (blue curves) and metal-enriched (red curves) SWCNT films before (dashed curves) and after (solid curves) annealing.}
\label{overview}
\end{center}
\end{figure}
%%%%%%%%%%

%We separated semiconducting and metallic SWCNTs using the technique of density gradient ultracentrifugation (DGU).\cite{ArnoldetAl05NL,ArnoldetAl06NN,YanagietAl08APE,HarozetAl10ACS,HarozetAl12JACS,HarozetAl13NS}  %The degree of separation is estimated to be 95\%.
As shown in Figure 1a, different coloration occurs between the metal- and semiconductor-enriched solutions due to the subtraction of different colors by distinct interband transitions.\cite{HarozetAl12JACS}  %We combined THz time-domain spectroscopy, Fourier transform infrared spectroscopy (FTIR), and NIR-VIS-UV spectroscopy to obtain broadband spectra (see Methods).
Figure 1b presents the attenuation, $-\log_{10}T$, where $T$ is the transmittance, in the NIR-VIS-UV range for solution and film samples of SWCNTs.  The M$_{11}$ and M$_{22}$ (S$_{22}$, S$_{33}$, and S$_{44}$) interband transitions are absent in the semiconductor (metal) case.  The semiconductor-enriched solution (blue solid line) exhibits sharp features, coming from different ($n$,$m$) species, which are not resolved in the corresponding film (blue dashed line).   This is expected because the nanotubes are individually suspended in the solution while they are bundled in the film, which broadens interband absorption peaks.  However, interestingly, in the metallic case, bundling-induced broadening seems non-existent, i.e., the M$_{11}$ and M$_{22}$ peaks maintain their shapes between the solution (red solid curve) and film (red dashed curve) forms.

Figure 1c shows the effect of annealing on the attenuation spectra of semiconductor- and metal-enriched films from the mid-infrared (MIR) to the UV, again showing the robustness of the interband features of metallic SWCNTs.  The films were vacuum annealed (see Section \ref{methods} for the annealing conditions used) to remove the adsorbed molecules that act as dopants to provide carriers.\cite{ZahabetAl00PRB,KangetAl05Nanotech}  The removal of carriers results in an enhancement of the S$_{11}$ peak as well as a small ($\sim$20~meV) red shift of both the S$_{11}$ and S$_{22}$ peaks.  We interpret these annealing-induced spectral changes in the semiconducting film in terms of Pauli blocking and bandgap renormalization; namely, before annealing, carriers occupy some near-band-edge states, which not only prevent interband transitions but also modify the bandgap {\it via} many-body interactions.  On the contrary, there are basically no doping effects in the metallic film, except the change seen in the S$_{11}$ region due to a small amount (estimated to be $\sim$5\%) of residual semiconducting SWCNTs. %The degree of separation is estimated as 95\%.
%Vacuum annealing also led to a 24~meV redshift in the S$_{11}$ transition and a 20~meV redshift in the S$_{22}$ peak.

%%%%% Fig. 2 %%%%%
\begin{figure}
\begin{center}
\includegraphics[scale=0.56]{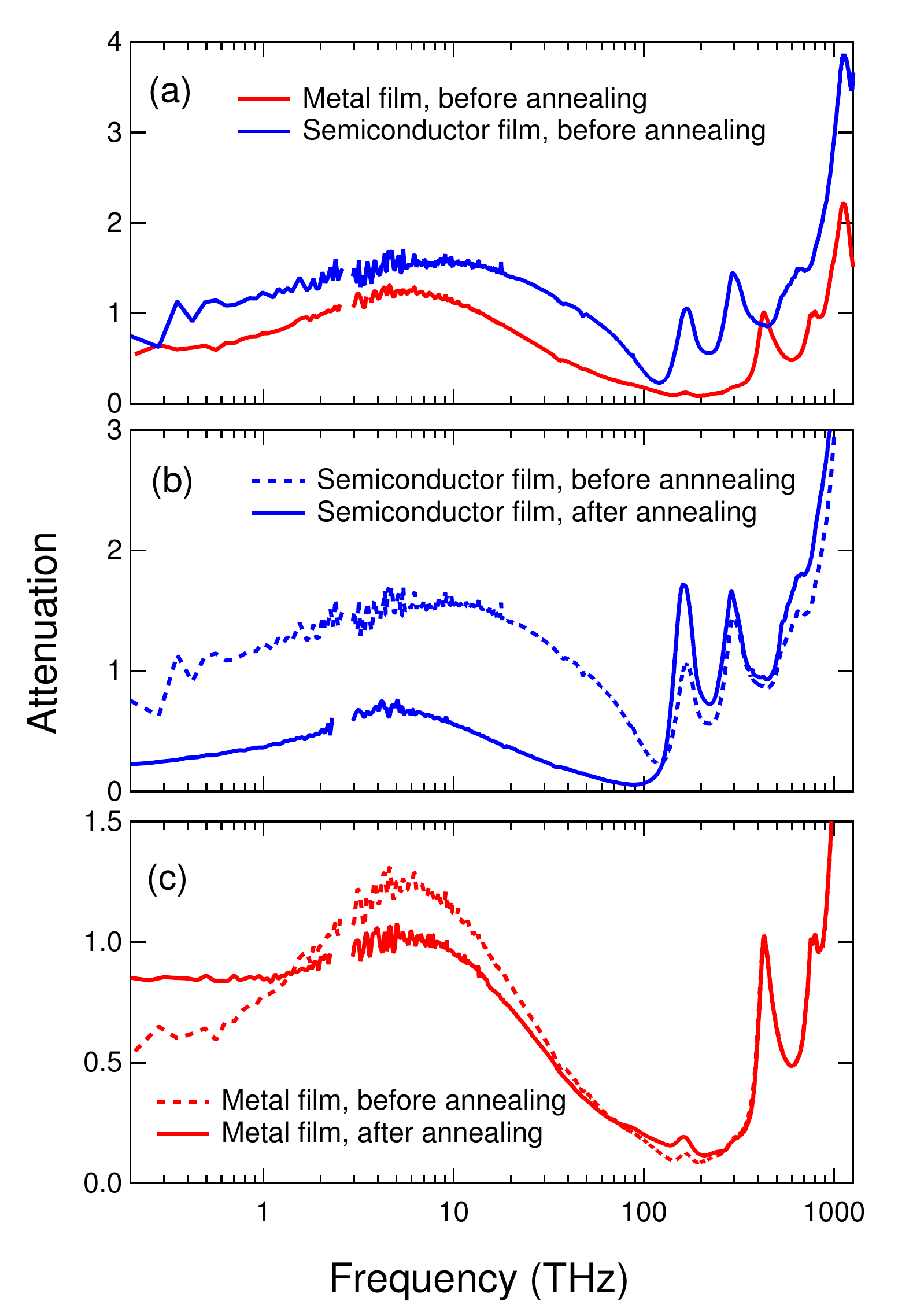}
\caption{THz-to-UV attenuation spectra of semiconductor- and metal-enriched SWCNT films at room temperature before and after annealing.  (a)~Semiconductor-enriched (blue) and metal-enriched (red) SWCNT films before annealing.  (b)~Semiconductor-enriched SWCNT film before (dashed) and after (solid) annealing.  (c)~Metal-enriched SWCNT film before (dashed) and after (solid) annealing.}
\label{anneal}
\end{center}
\end{figure}
%%%%%%%%%%
%%%%% Fig. 3 %%%%%
\begin{figure*}
\begin{center}
\includegraphics[scale=0.675]{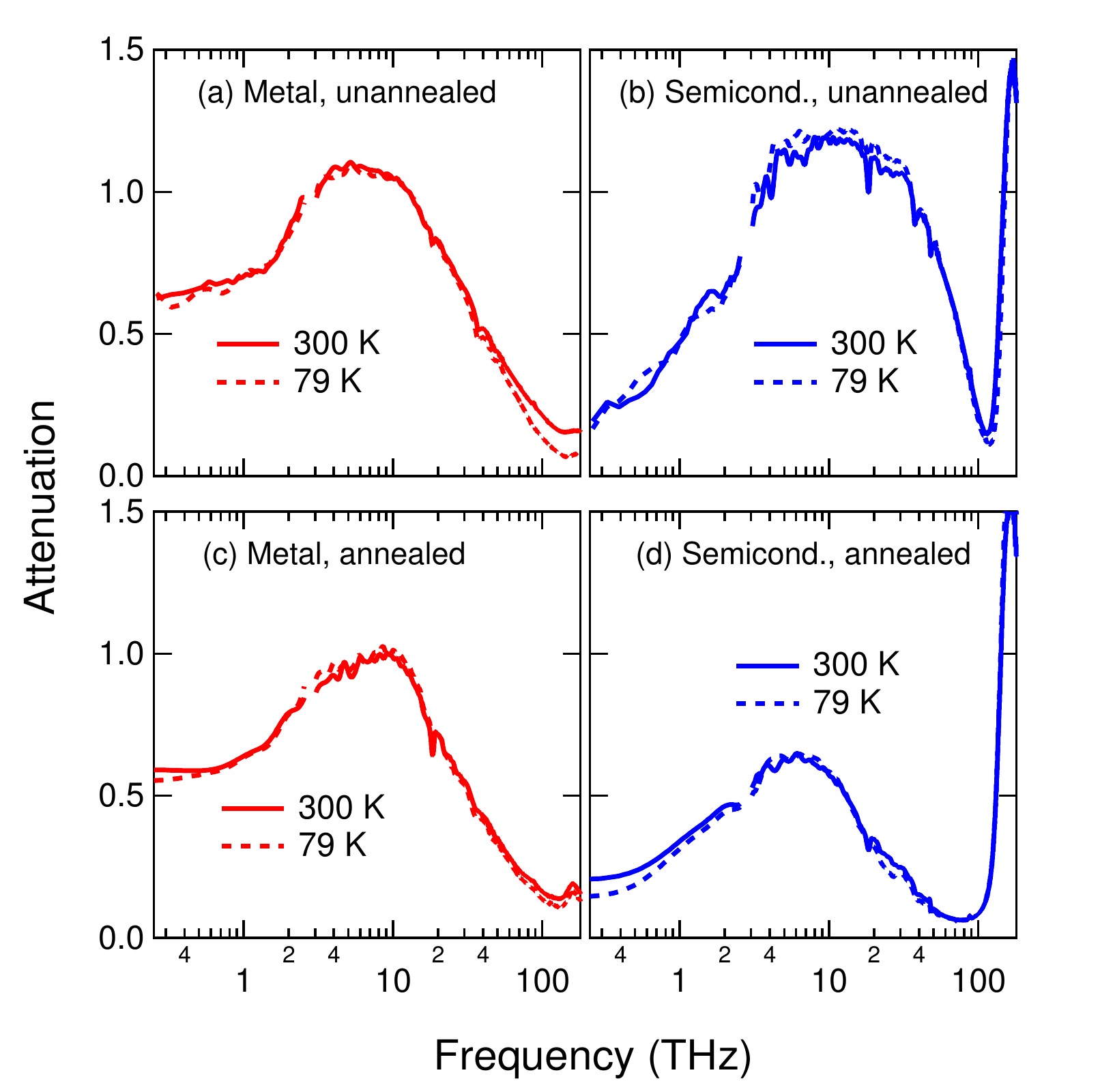}
\caption{Temperature dependence of the THz peak in (a)~metal-enriched SWCNT film before annealing, (b)~semiconductor-enriched SWCNT film before annealing, (c)~metal-enriched SWCNT film after annealing, and (d)~semiconductor-enriched film after annealing.}
\label{T-dep}
\end{center}
\end{figure*}
%%%%%%%%%%

Attenuation spectra of the semiconductor- and metal-enriched SWCNT films at room temperature in a very broad spectral range, from the THz to the UV, are shown in Figure 2a, demonstrating the existence of a broad and pronounced THz peak {\em in both types of films}; the peak is slightly smaller in width and lower in frequency in the metallic film.  Annealing effects are strikingly different between the two types of films:  in the semiconducting film (Figure 2b), the THz peak is significantly suppressed by carrier removal, whereas in the metallic film (Figure 2c), the THz peak exhibits a small decrease in intensity at its peak position but there is an increase in absorption at the lowest frequency side of the spectrum toward DC.  Note that the sharp features above 100~THz are interband transitions (S$_{11}$, S$_{22}$, S$_{33}$, M$_{11}$, and M$_{22}$), whose behaviors upon annealing are discussed in the last paragraph.  Furthermore, we performed temperature-dependent spectroscopy of this broad THz peak down to 79~K, as shown in Figure 3.  There is very little spectral change between 300~K and 79~K for both before (Figures 3a and 3b) and after (Figures 3c and 3d) annealing in both metallic (Figures 3a and 3c) and semiconducting (Figures 3b and 3d) nanotubes.

\section{Discussion}

Based on these observations, we can rule out the curvature-induced gap interpretation for the appearance of the THz peak.  There are three lines of argument that strongly support this conclusion.  First, interband transitions for curvature-induced gaps should exist only in metallic samples.  However, the THz peak clearly appears in the semiconducting film with a strength comparable to that in the metallic film.  Second, an optical excitation across a $\sim$20~meV gap should be extremely sensitive to carrier doping due to Pauli blocking.  The nanotubes need to be almost intrinsic in order to make the curvature-gap transition dominate the low-energy excitation.  Our data show the opposite: doping semiconducting nanotubes {\em enhances} the THz peak.  In the metallic film (which contains non-armchair metallic nanotubes), on the other hand, doping has only small influence.  Third, a very weak temperature dependence was observed for the THz peak.  A curvature-induced gap is calculated to be 0-20~meV (0 for armchair and 20~meV for zigzag tubes) for the metallic nanotubes used in this study ($d_t$ $\sim$ 1.5~nm),\cite{HarozetAl13NS} which is comparable with or much less than the room temperature thermal energy (26~meV).  Therefore, such interband absorption is expected to be very sensitive to the thermal carrier distribution, in contradiction with our observation.  %Even after vacuum anneal, THz peak of metal enriched film still shows very small change when be cooled down to 79~K. It clearly shows THz peak is not dominated by interband transition of curvature induced gap or other small gaps. But the weak temperature dependence feature is consistent with plasmon resonance model. Therefore, we conclude that the curvature-induced gap mechanism is disproved by our experiments, while plasmon resonance and free carrier response are supported.

%%%%% Fig. 4 %%%%%
\begin{figure*}
\begin{center}
\includegraphics[scale=0.7]{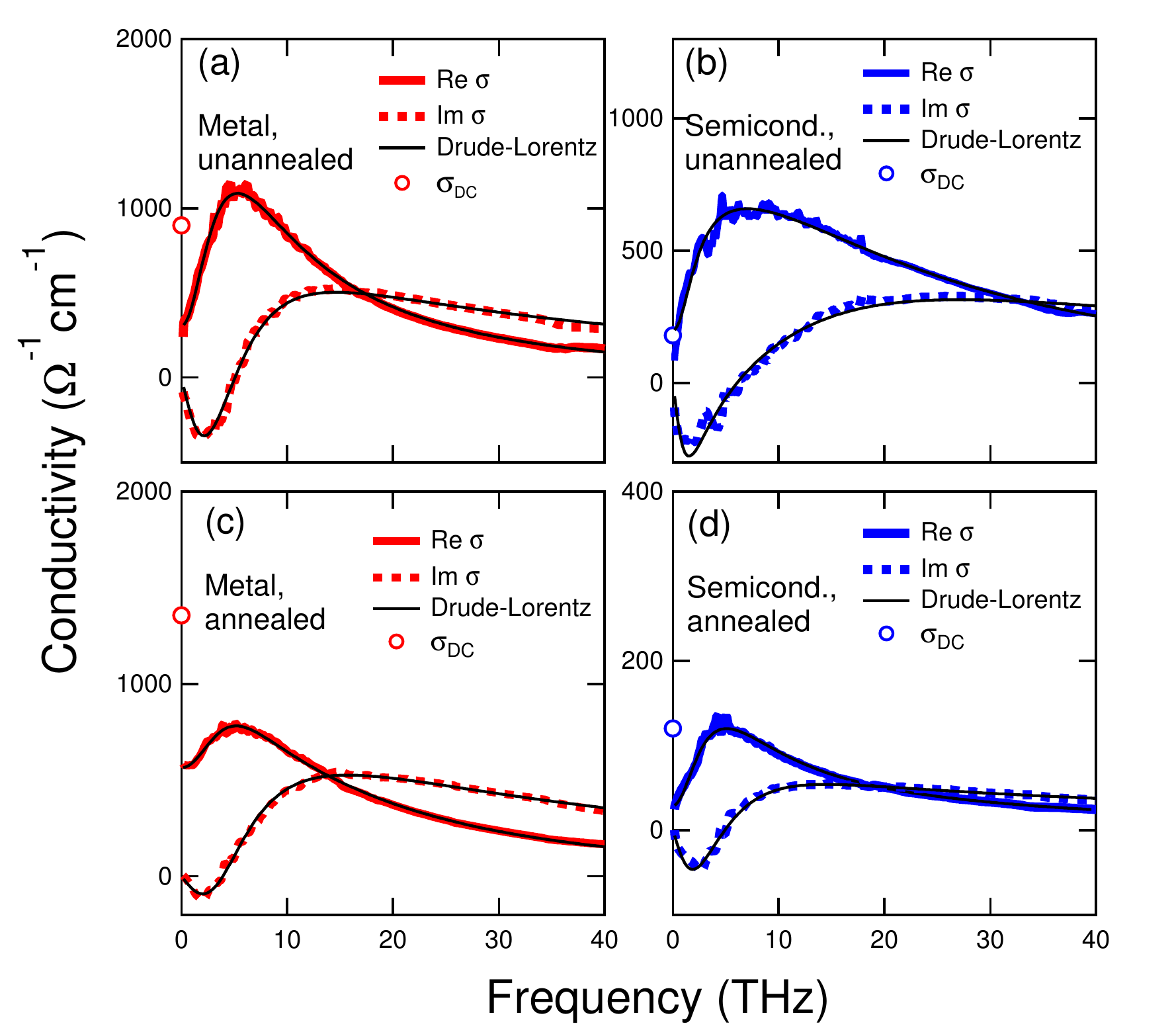}
\caption{DC and optical conductivity of semiconductor- and metal-enriched SWCNT films at room temperature before and after annealing. Real (solid lines) and imaginary (dashed lines) part of optical conductivity can be described by taking into account both the Drude-like free electron response and the quasi-1D plasmon resonance using eq 1.  Fitting curves are shown as black solid lines.  Open circles show the DC conductivity (located at zero frequency) obtained by DC transport measurements.}
\label{DC-THz}
\end{center}
\end{figure*}
%%%%%%%%%%

Our observations are qualitatively consistent with the plasmon resonance interpretation.  In particular, the THz peak is observed both in metallic and doped semiconducting nanotubes, its intensity is enhanced by doping in semiconducting SWCNTs, and there is no temperature dependence.  Also, strong polarization dependence has been universally confirmed by many authors, with no resonance expected for polarization perpendicular to the nanotube axis.\cite{JeonetAl02APL,AkimaetAl06AM,RenetAl13PRB}   

To gain more quantitative understanding via comparison with theoretical models, we performed Kramers-Kronig (KK) transformation on our broadband transmission spectra $T(\omega)$ and obtained the phase spectra $\phi_{\rm t}(\omega)$.\cite{BorondicsetAl06PRB,PekkerKamars11PRB,DresselGruener02Book} 
We first interpolated the $T(\omega)$ data in the region between the THz and IR and then applied four-pass binomial smoothing to prevent spikes.  On the UV side, we extrapolated $T(\omega)$ by using $1-A\omega^{-n}$, where $A$ is a real constant and $n$ is around $1$.  We chose $A$ and $n$ to make the extrapolation as continuous and smooth as possible.  On the THz side, we did linear extrapolation of $T(\omega)$ to the zero frequency.  Finally, by using
%%%
\begin{equation}
\phi_{\rm t}(\omega) = { -{\omega \over \pi}\int {\ln(T(\omega'))-\ln(T(\omega)) \over \omega'^2-\omega^2}d\omega'},
\end{equation}
%%%
we were able to obtain the phase spectrum, $\phi_{\rm t}(\omega)$.  With $\phi_{\rm t}(\omega)$ and $T(\omega)$, we obtained the transmission coefficient, $t(\omega)=\sqrt{T(\omega)} \exp(i\phi_{\rm t}(\omega))$.

We then converted the transmission coefficient, $t(\omega)$, into the complex optical conductivity, $\sigma(\omega)$, through $t(\omega)= (1+n_{\rm sub}) / (1+n_{\rm sub}+Z_0 \sigma(\omega) d)$ under the thin-film approximation,\cite{NussOrenstein98THz}  which is valid from DC to the MIR in our case.  Here, $Z_0$ = 377~$\Omega$ is the vacuum impedance, $d$ is the film thickness, and $n_{\rm sub}$ is the substrate refractive index.  The real and imaginary parts of $\sigma(\omega)$ are plotted in Figure 4 for both films for both before and after annealing.  One immediately notices the asymmetric lineshape of the THz peak: the conductivity decays more slowly on the high frequency side.  It is also noticeable that each trace has a finite conductivity in the DC limit, whose trend agrees with our DC transport measurements.  Annealing results in changes in the DC values, also consistent with the $\sigma(\omega)$ results.  The  quantitative difference is likely due to the inhomogeneity of the current distribution in the DC measurements; THz radiation acts as a local probe, which is not sensitive to macroscopic inhomogeneity.

The observed asymmetric lineshape cannot be reproduced by a single Lorentzian representing the plasmon resonance; it also cannot be fitted by the Drude-Smith model representing carrier localization.\cite{Smith01PRB}  The finite DC conductance proves the existence of macroscopic percolating conduction channels.  Namely, our film cannot be treated as a network of isolated nanotubes (or nanotube bundles) surrounded by air; a correct model has to take into account the intertube transport responsible for the finite macroscopic DC conductance.  Therefore, we take into account two effects in our model to describe the experimental optical conductivity: (1) plasmon resonance, which represents the confined collective motion of carriers in the tube-length scale, and (2) Drude-like free carrier response, which describes the intertube transport and percolating channels in the macroscopic scale. The fitting equation we use is
%%% Eq 1 %%%
\begin{equation}
\sigma(\omega) = {i\sigma_{\rm plasmon} \omega\gamma_{\rm plasmon} \over \omega^2 - \omega_0^2 +  i\omega\gamma_{\rm plasmon}} + {i\sigma_{\rm Drude} \gamma_{\rm Drude} \over \omega + i\gamma_{\rm Drude}},
\label{eq1}
\end{equation}
%%%%%%%%%
where the first (second) term represents the contribution from the plasmon resonance (the Drude-like free electron response), $\sigma_{\rm plasmon}$ is the plasmon conductivity at the resonance frequency ($\omega = \omega_0$), and $\sigma_{\rm Drude}$ is the Drude conductivity in the DC limit. $\gamma_{\rm plasmon}$ and $\gamma_{\rm Drude}$ are phenomenological scattering rates for the plasmon and free election response, respectively. This equation describes the experimental traces well, as shown in Figure 4.  The extracted fitting parameters are listed in Table 1.

%\begin{widetext}
%%% Table 1 %%%
\begin{table*}
  \caption{The parameters extracted through fitting the THz optical conductivity spectra with eq~1.  $\sigma_{\rm plasmon}$ is the plasmon conductivity at the resonance frequency ($\omega_0$), $\sigma_{\rm Drude}$ is the Drude conductivity in the DC limit,  $\gamma_{\rm plasmon}$ is the plasmon scattering rate, $\gamma_{\rm Drude}$ is the free electron scattering rate, and $\sigma_{\rm DC}$ is the DC conductivity determined through transport measurements,}
  \label{table1}
  \begin{tabular}{l|cccc}
    \hline
    \hline
     & Semiconductor & Metal & Semiconductor & Metal\\
     & & & annealed & annealed \\
    \hline
    $\sigma_{\rm plasmon}$ ($\times$10$^{2}$~$\Omega^{-1}$cm$^{-1}$) & 5.0$\pm$0.1 & 7.9$\pm$0.3 & 0.91$\pm$0.1 & 2.5$\pm$0.1 \\
    $\sigma_{\rm Drude}$ ($\times$10$^{2}$~$\Omega^{-1}$cm$^{-1}$) & 1.3$\pm$0.1 & 2.8$\pm$0.1 & 0.28$\pm$0.05 & 5.6$\pm$0.1\\
    $\omega_0 /$2$\pi$ (THz) & 6.6$\pm$0.2 & 5.5$\pm$0.1 & 5.1$\pm$0.1 & 5.5$\pm$0.1\\
    $\gamma_{\rm  plasmon}$ (THz) & 152$\pm$10 & 72$\pm$5 & 73$\pm$5 & 66$\pm$5 \\
    $\gamma_{\rm  Drude}$ (THz) & 300$\pm$20 & 142$\pm$10 & 220$\pm$20 & 140$\pm$10 \\
    $\sigma_{\rm DC}$ ($\times$10$^{2}$~$\Omega^{-1}$cm$^{-1}$) & 1.8$\pm$0.1 & 8.9$\pm$0.1 & 1.2$\pm$0.1 & 14.0$\pm$0.1\\
    \hline
    \hline
  \end{tabular}
\end{table*}
%%%%%%%%
%\end{widetext}

The plasmon resonance frequency $\omega_0$ for a SWCNT with length $l$ is given by $\omega_0$ = $v_q (\pi/l)$, where $v_q$ is the mode velocity, which is equal to $4v_{\rm F} \left [ {e^2 \over \sqrt{3} \pi \varepsilon a_{\rm C-C} \gamma_0} K_0 ({\pi d_t \over 2l}) I_0 ({\pi d_t \over 2l})\right ]^{1/2}$, where $v_{\rm F}$ is the Fermi velocity, $\varepsilon$ is the static dielectric constant, and $K_0$ and $I_0$ are the first and second type, respectively, of the modified Bessel function.\cite{NakanishiAndo09JPSJ}  Using $v_{\rm F}$ = 10$^8$~cm/s,  $\varepsilon$ = 2.5,\cite{TaftPhilipp65PR} $d_{\rm t}$ = 1.5~nm, $a_{\rm C-C}$ = 0.142~nm, and $\gamma_0$ $\sim$ 3.2~eV, together with the measured value of $\omega_0/2\pi$ = 5.5~THz (for metal-enriched films before and after annealing), we obtain $l$ = 340~nm, very close to the typical tube length in our sample (300~nm).  This confirms that the THz conductivity peak is due to plasmon resonance.  In the semiconductor case, the plasmon resonance occurs at a slightly higher frequency, while the scattering rate is almost as twice large as that in the metal case.  After annealing, the residual THz peak in the semiconductor film has $\omega_0$ and $\gamma_{\rm plasmon}$ values that are very similar to the metallic film.  We thus interpret the residual peak as mainly arising from the residual metallic tubes in the semiconductor enriched films.  The ratio between the carrier densities of the annealed semiconductor and metal films, which can be extracted by spectrally integrating the real part of the conductivity, is $0.12$. This number should be considered to be the upper limit of the residual metallic tubes in the semiconductor-enriched films since it also includes defects-induced doping, which cannot be removed by annealing.  In the annealed metal film, the Drude term dominates the low-frequency response ($\sigma_{\rm Drude}$ $>$ $\sigma_{\rm plasmon}$), unlike the other three films where $\sigma_{\rm Drude}$ $<$ $\sigma_{\rm plasmon}$.  The result indicates that intertube carrier migration is enhanced by the removal of trapped molecules and the annealing-induced nanotube distortion around the junction, which increases the contact area between nanotubes.  In addition, the small $\sigma_{\rm Drude}$ in the annealed semiconductor film can be understood as the  lack of macroscopic percolating channels for the residual metallic nanotubes.

\section{Conclusions}

In summary, we performed transmission spectroscopy over a wide range (from the THz to the UV) as well as DC transport measurements of semiconductor- and metal-enriched SWCNT samples.  Our experimental results show that the broad THz peak originates from a plasmon resonance in both the metallic and doped semiconducting carbon nanotubes rather than the interband excitation of the curvature-induced gap in non-armchair metallic nanotubes.  Intraband free electron response also contributes to the low-energy excitation spectrum, especially in the metal-enriched film after annealing.  Our studies provide fundamental insight into the low-energy excitation in SWCNTs, while the broadband spectroscopy of semiconducting and metallic type-separated nanotube samples also provide basic knowledge useful for emerging applications of SWCNTs in plasmonics and optoelectronics in the technologically important THz frequency range.  Future studies on films of SWCNTs enriched in single chiralities should provide further information on the chirality dependence of low-energy excitations in SWCNTs.  In particular, samples of single-chirality non-armchair metallic nanotubes, such as (7,4) nanotubes, will be useful for elucidating the nature of interband absorption across the curvature-induced gaps.

%% classes.  Rather than use \section, an appropriate macro is
%% provided that will always work.
%%%%%%%%%%%%%%%%%%%%%%%%%%%%%%%%%%%%%%%%%%%%%%%%%%%%%%%%%%%%%%%%%%%%%
\section*{Acknowledgements}

This work was supported by the Department of Energy (through Grant No.~DE-FG02-06ER46308), the National Science Foundation (through Grants No.~OISE-0968405 and EEC-0540832), and the Robert A. Welch Foundation (through
Grant No.~C-1509).

%\pagebreak

%\bibliography{jun}

\end{document}